\documentclass[twocolumn,showpacs,preprintnumbers,amsmath,amssymb]{revtex4}

\usepackage{wasysym}
\usepackage{epsfig}
\usepackage{graphicx}
\usepackage{dcolumn}
\usepackage{bm}
\usepackage{verbatim}

\newcommand{\refeq}[1]{(\ref{#1})}

\newcommand{\om}{\omega}

\newcommand{\be}{\begin{equation}}
\newcommand{\ee}{\end{equation}}
\newcommand{\bea}{\begin{eqnarray}}
\newcommand{\eea}{\end{eqnarray}}
\newcommand{\eps}{\epsilon}

\newcommand{\LL}{\mathcal L}

\begin{document}

\preprint{APS/123-QED}

\title{Gravitational Redshift, Equivalence Principle, and Matter Waves}
\author{Michael A.~Hohensee}
\author{Brian Estey}
\author{Francisco Monsalve}
\author{Geena Kim}
\author{Pei-Chen Kuan}
\author{Shau-Yu Lan}
\affiliation{Department of Physics, 366 Le Conte Hall MS 7300, University of California, Berkeley, California 94720, USA}
\author{Nan Yu}
\affiliation{Jet Propulsion Laboratory, M/S 298-100, 4800 Oak Grove Dr., Pasadena, CA 91109}
\author{Achim Peters}
\affiliation{Humboldt-Universit\"at zu Berlin, Hausvogteiplatz 5-7, 10117 Berlin, Germany}
\author{Steven Chu}
\affiliation{US Department of Energy, 1000 Independence Avenue SW, Washington, District of Columbia 20585, USA}
\affiliation{Department of Physics, 366 Le Conte Hall MS 7300, University of California, Berkeley, California 94720, USA}
\affiliation{Lawrence Berkeley National Laboratory, One Cyclotron Road, Berkeley, California 94720, USA}
\author{Holger M\"uller}
\email{hm@berkeley.edu}
\affiliation{Department of Physics, 366 Le Conte Hall MS 7300, University of California, Berkeley, California 94720, USA}
\affiliation{Lawrence Berkeley National Laboratory, One Cyclotron Road, Berkeley, California 94720, USA}

\date{\today}

\begin{abstract}
We review matter wave and clock comparison tests of the gravitational redshift. To elucidate their relationship to tests of the universality of free  fall (UFF), we define scenarios wherein redshift violations are coupled to violations of UFF (``type II"), or independent of UFF violations (``type III"), respectively. Clock comparisons and atom interferometers are sensitive to similar effects in type II and precisely the same effects in type III scenarios, although type III violations remain poorly constrained. Finally, we describe the ``Geodesic Explorer," a conceptual spaceborne atom interferometer that will test the gravitational redshift with an accuracy 5 orders of magnitude better than current terrestrial redshift experiments for type II scenarios and 12 orders of magnitude better for type III.
\end{abstract}

\maketitle


According to the Einstein Equivalence Principle, the laws of Special Relativity hold locally in any freely falling frame. As a result, lowering a clock into a gravitational potential by $\Delta U$ will slow it down by a factor of $1+\Delta U/c^2$, where $c$ is the velocity of light. This gravitational redshift was the first  consequence of General Relativity described by Einstein \cite{Einstein1911}, and its measurement continues to be fundamental to our confidence in the theory. Clock comparison tests have measured the redshift with an accuracy of $7\times 10^{-5}$ \cite{Vessot} (see Tables \ref{exptstable} and \ref{Nulltesttable}), while matter wave tests, in which redshift anomalies modify material particles' Compton frequencies, have reached accuracies of $7\times 10^{-9}$ \cite{redshift}. However, the principle of energy and momentum conservation intimately links the gravitational redshift to another consequence of the Einstein Equivalence Principle, the universality of free fall \cite{Nordtvedt1975,Haugan1978,TEGP,EEPremark}. The indirect bounds on redshift violations derived from such tests can be more stringent than the current direct bounds. Here, we review clock comparison and matter wave interference experiments, showing that modern redshift experiments can nevertheless bound anomalies that are presently poorly constrained. Using the Standard Model Extension (SME) \cite{KosteleckyTasson}, we present explicit lagrangians for redshift violations. Redshift violations that are independent of violations of UFF arise at post newtonian order 4 due to the nonlinearity of general relativity \cite{MikeCPT}. In addition, we note that the most commonly used model for redshift violation does not produce physically measurable effects, although these can be recovered in more subtle scenarios. We identify the similarities and differences between matter wave and clock comparison tests. 
We obtain direct bounds on several parameters of the SME \cite{KosteleckyTasson}. We also propose new experiments that could improve the current bounds on redshift violations to $10^{-14}$, beyond the accuracy of current UFF tests.

\begin{table*}
\caption{\label{exptstable} Selected measurements of the gravitational redshift. Source masses for the gravitational field included Earth ($\earth$), Saturn ($\saturn$), the Sun ($\astrosun$), and neutron stars (NS). ISS; international space station. See \cite{Hentschel} for a review of experiments before 1972. A result in parenthesis indicates the goal of a proposed experiment.}
\begin{tabular}{cccccc}\hline\hline
Ref. & Clock & Source & Separation & Comparison & Accuracy \\ \hline
\cite{HK1,HK2} & Cs clock & $\earth$ & 10\,km, aircraft & Transport & 0.1 \\
\cite{Cottam} & Inner-shell lines  & NS & $z=0.35$  & Photons & several \% \\
\cite{PoundRebkaSnyder} & $^{57}$Fe M\"ossbauer & $\earth$ & 22.5\,m, tower & Photons & $1\times 10^{-2}$ \\
\cite{Krisher1990} & Crystal oscillator & $\saturn$ & Orbit & Radio link & $1\times 10^{-2}$ \\
\cite{Colella,Bonse,Horne,Werner,Littrell} & Neutron matter waves & $\earth$ & $\sim$ cm, diffraction & Transport & 1\% \\
\cite{Krisher} & Crystal oscillator & $\astrosun$ & Orbit & Radio link & $5\times 10^{-3}$\\
\cite{Vessot} & H-maser & $\earth$ & $10^7\,$m, rocket & 2-way radio & $7\times 10^{-5}$ \\
\cite{redshift} & Cs matter waves & $\earth$ & 1.2\,mm, photon recoil & Transport & $7\times 10^{-9}$ \\
\cite{ACES}$^a$ & H and Cs clocks & $\earth$ & 350\,km, ISS & 2-way radio/optical & ($2\times 10^{-6}$) \\
\cite{EGE}$^a$ & Cs and optical & $\earth$ & Satellite & 2-way  radio & ($2.5\times 10^{-8}$) \\
\hline\hline
\end{tabular}\\
{\footnotesize $^a$ planned}
\end{table*}

\begin{table}
\caption{\label{Nulltesttable} Null tests. Two different types of clocks are compared in a laboratory on Earth, whose gravitational potential varies as a result of earth's orbit. The absence of a relative frequency change of the clocks is verified to within the stated accuracy.}
\begin{tabular}{cccc}\hline\hline
Ref. & Clock 1 & Clock 2 &  accuracy \\ \hline
\cite{Braxmaier} & Cavity & I$_2$ & $4\times 10^{-2}$ \\
\cite{Turneaure} & Cavity & H-maser & $1.7\times 10^{-2}$ \\
\cite{Godone} & Mg & Cs &  $7\times 10^{-4}$ \\
\cite{Tobar} & Cavity & H-maser & $(-2.7\pm 1.4)\times 10^{-4}$ \\
\cite{Bauch} & Cs & H-maser &  $2.1\times 10^{-5}$\\
\cite{Fortier} & $^{199}$Hg & Cs &  $(2.0\pm 3.5)\times 10^{-6}$ \\
\cite{Blatt} & Cs & $^{87}$Sr &  $3.5\times 10^{-6}$ \\
\cite{Ashby} & Cs & H-maser & $(0.1\pm1.4)\times 10^{-6}$ \\
\hline\hline
\end{tabular}
\end{table}

The proper time interval $d\tau$ in general relativity is given by \cite{MTW}
\be
cd\tau=\sqrt{-g_{\mu\nu}dx^\mu dx^\nu}.
\ee
This combines the gravitational redshift as function of location $r$ with special relativistic time dilation. The metric $g_{\mu\nu}$ (where $\mu,\nu=0,1,2,3$) encodes the properties of space-time and the $x^\mu$ are the coordinates. In particular, the proper times of two clocks at rest at different locations $x^\mu$ and $x'{}^\mu$ have a ratio
\be
\frac{d\tau}{d\tau'}=\sqrt{\frac{g_{00}}{g'{}_{00}}},
\ee
which is the expression for the gravitational redshift in general relativity. Since a freely falling particle seeks the path that maximizes its proper time, the gravitational redshift as given by $g_{00}$ is directly responsible for the gravitational force on a slowly moving test particle. For a particle moving radially in the Schwarzschild metric in Schwarzschild coordinates $t, r, \theta,\phi$, this expression reduces to
\bea\label{nonrel}
\tau=\int\sqrt{-g_{00}(r)-g_{rr}v^2}dt \approx \int \left(1+\frac{gz}{c^2} -\frac12\frac{v^2}{c^2}\right)dt
\eea
for a clock moving with velocity $v$ in radial direction. For taking the nonrelativistic limit, we dropped a constant term and denoted $z\ll r$ the vertical coordinate in the laboratory frame and $g$ the acceleration of free fall.

\begin{figure}
\centering
\epsfig{file=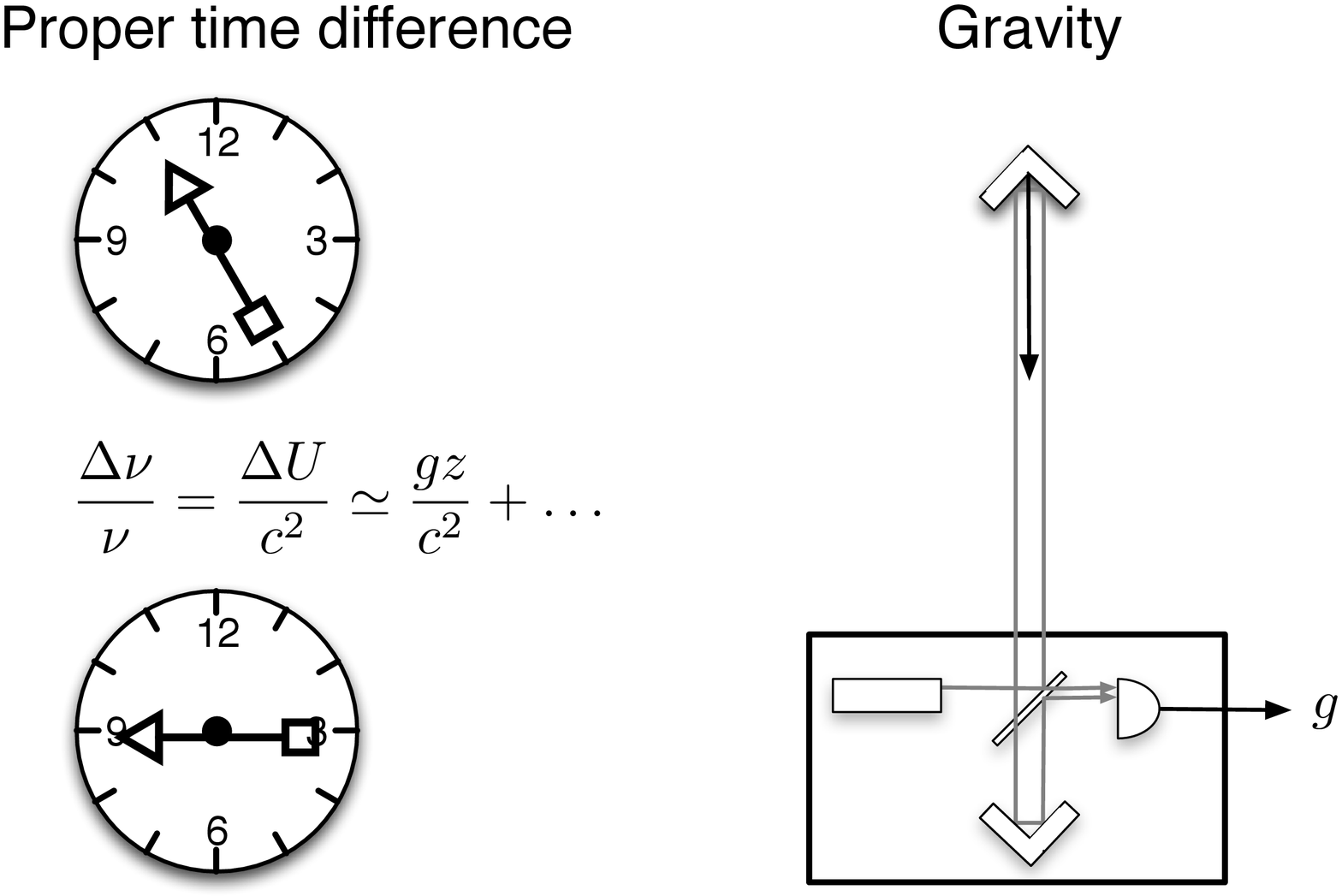,width=0.45\textwidth}
\caption{\label{GenericTest} Generic redshift experiment: the gravitational redshift is measured by comparing two clocks at different locations, and the gravitational potential difference $\Delta U$ between the two locations is measured by monitoring the geodesic trajectory between these locations. In the limit that the experiment is so small that the gravitational acceleration can be taken as a constant, this reduces to the measurement of the gravitational acceleration.}
\end{figure}

Figure \ref{GenericTest} shows a generic redshift measurement between two stationary clocks at different locations. In past experiments, exchange of electromagnetic signals ({\em e.g.}, in Ref. \cite{PoundRebkaSnyder}) or clock transport (in Ref. \cite{HK1,HK2}) has been applied for making the comparison. Note that the redshift is $\Delta U/c^2$, independent of the method used for the comparison.

In order to compare the measured redshift with the theoretical prediction, the gravitational potential difference between the two clocks has to be known. This knowledge is obtained by monitoring the geodesic trajectory of an object. In satellite redshift tests \cite{Vessot,ACES}, this may be accomplished by monitoring the trajectory of the craft itself. For a laboratory experiment, the inhomogeneity of the gravitational field can usually be ignored to first order, so that the measurement of the potential difference $\Delta U=gz+\mathcal O(z^2)$ reduces to a measurement of the clock's separation $z$ and the gravitational acceleration $g$. 

Since the geodesic trajectory maximizes the proper time, a change of the gravitational redshift might change the outcome of the measurement of the potential difference. If the outcome of both measurements change by the same amount, no anomaly will be detectable. It is therefore important to study the phenomenology of redshift violations, to see in which cases a meaningful ({\em i.e.}, measurable) anomaly is obtained.



\section{Phenomenology of redshift violations}

\subsection{Type I}

At first glance, we might expect to obtain an anomalous redshift from attaching a factor to the redshift term in Eq. (\ref{nonrel}):
\be\label{betalag}
\LL=mc^2\left[(1+\beta)\frac{gz}{c^2}-\frac 12 \frac{\dot z}{c^2}\right].
\ee
Here the dimensionless constant $\beta$, taken to be constant for all objects, vanishes if general relativity is valid. This Lagrangian can be shown to yield no measurable anomalies: Rescaling the units of time and mass according to $t\rightarrow (1-\beta/2)t$ and $m\rightarrow (1+\beta)m$ absorbs the $\beta$ into an unmeasurable redefinition of coordinates. To illustrate this important fact, consider the generic test experiment shown in Fig. \ref{GenericTest}. If  $\beta$ is nonzero, the redshift measurement will find a redshift given by $(1+\beta) gz/c^2$. However, the measurement of the gravitational potential difference will now yield $(1+\beta) gz$. Since the results of both experiments are shifted in the same way, the anomaly will cancel upon comparing the measured and predicted redshift. Thus, ``type I violations" do not constitute a deviation from standard physics and are thus not measurable in any experiment.

\subsection{Type II}

If the parameter $\beta=\beta^{\rm T}$ varies with the composition of the clock or test particle T, measurable effects are obtained. The result will depend on the difference $\beta^{\rm T}-\beta^{\rm U}$ of two particles T and U. Tests of the gravitational redshift and of UFF will both lead to nonzero results. However, since UFF has been experimentally confirmed to an accuracy of $10^{-13}$ \cite{Schlamminger2008,Adelberger2009}, redshift tests must be extremely accurate to yield new constraints.

The situation for clock comparisons is a little bit better if the parameter $\beta^{\rm T}$ is nonzero for electromagnetic masses only: Electromagnetic interactions contribute all of the signal in atomic clocks such as hydrogen masers or cesium clocks, but only a fraction of a particle's rest mass. Thus, for a given violation, signals in redshift tests are larger than in UFF tests. However, given the current accuracy of UFF tests, redshift experiments with atomic clocks would need to reach an accuracy of $2\times 10^{-7}$ before they can become competitive even in this (arguably specialized) situation, see p. 29 in Ref. \cite{TEGP} and \cite{Nordtvedt1975,Haugan1978}.

A famous, though unproven, conjecture by Schiff \cite{Schiff,Dicke} states that in any consistent theory of gravitation, all violations of the Einstein Equivalence Principle are of type II. Theories that may evade this conjecture will be discussed below.


\subsection{Type III}

We define type III violations as the case where the gravitational redshift is changed by a factor of $1+\beta_{\rm III}$ while the acceleration of free fall is universal. In Sec. (\ref{typeIII}), it will turn out that the leading order type III effect depends on the velocity $v$ of the experiment along the gradient of the gravitational potential. Hence, $\beta_{\rm III}$ is a function of that velocity. Specialized models for type III violations have been proposed, {\em e.g.,} based on nonminimal coupling \cite{Coley,Ni1977,Accioly}. A combination of gravitoscalar and gravitovector fields produce models exhibiting type II violation, although in special cases their effect on normal matter can cancel \cite{Nieto}.  The remaining type II effect would then be apparent in comparisons between electromagnetic binding energy and normal matter.  This special situation is similar to type III, because electromagnetism contributes only a small part to particle rest masses. Unfortunately, the detailed experimental consequences of these models have not yet been worked out. Below, we will derive a detailed model for type III violations.  Type III scenarios are arguably the most interesting ones for redshift tests, as they lead to a violation of the Einstein Equivalence Principle that cannot be detected by tests of UFF.

\subsubsection*{Energy and momentum conservation}

It has been shown \cite{Nordtvedt1975,Haugan1978,TEGP} that redshift violations not ruled out by UFF tests require violation of conventional energy conservation. Such a scenario is not as counterintuitive as it might at first appear: A complete model for type 3 effects would explain the violation by the interaction of the system with an additional ``fifth force" field. Then, the conserved energy includes the additional field, leading to an apparent non-conservation if the additional field is not accounted for. Another possibility is that energy is conserved in a nonlocal fashion. In both cases, a type III redshift violation could be obtained while globally, energy and momentum are conserved.

In the explicit scenario presented in Sec. \ref{typeIII}, UFF is (locally) preserved while the gravitational redshift is anomalous. Energy and momentum are conserved (as derived from the Lagrangian in the usual way), but include a contribution of the redshift-violating field.

Another example is the theory put forward by \cite{Nieto}. There are two anomalous gravitational couplings: First, a scalar field that couples to the stress-energy of matter or other fields {\em e.g.,} electromagnetism, as well as to a conserved charge like baryon number. Second, a vector field that  couples only to the conserved current. Because the vector coupling can lead to a repulsive force for normal matter (rather than antimatter), whereas the scalar coupling leads to attraction, complete cancelation of the forces is possible. Thus, UFF is preserved for normal matter but the scalar coupling could still violate the redshift by its effect on the mass-energy of a system. There is no contradiction with the argument given in \cite{Nordtvedt1975}, because the free-fall rates of different quantum states which were considered there would be affected only by the scalar coupling \cite{Krisher1995}.



\section{Clock comparison tests}

Absolute redshift measurements (Tab. \ref{exptstable}) compare clocks at locations whose gravitational potential differs by $\Delta U$ and test whether their relative frequency change $\Delta \nu/\nu$ satisfies $\Delta\nu/\nu=\Delta U/c^2$. These tests have so far used various types of clocks ranging from nuclear (M\"ossbauer) transitions to mechanical vibrations. They have also used different source masses for the gravitational potential and the clocks have been compared using both clock transport and photon signals. In general relativity, the gravitational redshift is unaffected by these variations in experimental procedure. However, the variety of tests is useful in ruling out different scenarios for violations.

For example, between 1959 and 1964, Pound, Rebka and Snider \cite{PoundRebkaSnyder} studied M\"ossbauer sources vertically separated by about $h=22.5$\,m in a tower at Harvard University. The upper source emits 14\,keV gamma quanta. When they are detected by the lower M\"ossbauer absorber, they will be blue-shifted by $hg/c^2 \sim 2.5\times 10^{-15}$. This shift is measured by moving the lower absorber at a velocity such that the gravitational redshift is compensated for in the frame of reference of the absorber. This occurs at a velocity of $hg/c$.


Vessot {\em et al.} \cite{Vessot} conducted the most precise absolute redshift measurement so far. In 1976, they launched Gravity Probe A (GP-A), a hydrogen maser on board a Scout rocket, to a height of 10,000\,km. The high elevation increases the redshift, but comes at the cost that now the clock is moving. Therefore, the special relativistic time dilation of the moving clock has to be corrected for. This is done by measuring the trajectory of the craft. Monitoring the trajectory rather than predicting it from some theoretical model provides a high degree of reliability, given that the trajectory might be modified by effects from standard physics ({\em e.g.}, friction) or nonstandard physics such as violations of UFF. Specifically, the result of Gravity Probe A was \cite{Vessot}
\bea
\frac{\Delta f}{f}&=&(1+\beta)\left(\frac{U_s-U_e}{c^2}-\frac{|\vec v_s-\vec v_e|^2}{2c^2}\right),\nonumber \\
\beta & = & (2.5\pm 63)\times 10^{-6},
\eea
where $\beta$ is $\beta_{\rm II}$ or $\beta_{\rm III}$ in type II or III scenarios, respectively. The first term in the bracket is the gravitational redshift due to the difference of the gravitational potentials of the transmitter and receiver, and the second term is special relativistic time dilation due to the relative motion (the linear Doppler effect is taken out by use of a two-way radio link \cite{Vessot}).

Relative redshift measurements (Tab. \ref{Nulltesttable}) use two different clocks at (nearly) the same location. The tests verify that no variation in their frequency ratio is caused by a modulation of their common gravitational potential. In practice, the highest precision tests use two clocks on Earth. The ellipticity of Earth's orbit means that the Sun's gravitational potential on Earth $\Delta U/c^2$ is modulated with an amplitude of $3.3\times 10^{-10}$ and a one year period. A smaller daily modulation due to Earth's rotation also exists.


\section{Matter wave tests}
\label{matterwavetests}

The first experiments to detect the influence of gravity on the phase of matter waves were performed by Colella and others \cite{Colella,Bonse,Horne,Werner,Littrell}. They reached an agreement with the expected value of better than 1\%. The basis for matter wave tests of the gravitational redshift is the action of a point particle in general relativity is given by \cite{MTW}, pp.315-324
\be
S= mc^2\int d\tau,
\ee
In Feynman's path integral formulation \cite{FeynmanHibbs} of quantum mechanics, the state $\psi(t_A,\vec x_A)$ is given by the path integral
\be
\psi(t_A,\vec x_A)=\int d^3 x_B \psi(t_B,\vec x_B)=\int_A^B D q(t) e^{\frac i\hbar\int mc^2 d\tau}.
\ee
This shows that the evolution of a quantum state is given by the gravitational redshift and time dilation as encoded in the proper time $\tau$.

In the semiclassical limit, the path integral reduces to an ordinary integral over the classical, geodesic path. This leads to the expression \cite{Borde5D}
\be\label{Comptonphase}
\phi=\frac i\hbar \int mc^2d\tau=\int \om_Cd\tau,
\ee
where $\om_C$ is the Compton frequency.

\subsection{``Pound-Rebka like" tests: stationary atoms in optical lattices}

Bloch oscillations of cold atoms in optical lattices \cite{Peik} are a well known effect in atomic physics. In the experiments of interest here, atoms are held in a coherent superposition of states located above one another with a separation of $\lambda/2$. 
In such a setup, a periodic modulation of the atom's quasimomentum is observed. We show that this modulation can be interpreted by the beat frequency between redshifted Compton frequencies of atoms located in adjacent potential minima. A limit on redshift violations can therefore be derived from experiments.
If we evaluate Eq. (\ref{Comptonphase}) for atomic states located in adjacent lattice sites at heights $z=n\lambda/2$ and $(n+1)\lambda/2$, we obtain a difference
\be
\om_B=(1+\beta)\frac\lambda 2 \frac{g}{c^2}\frac{mc^2}{\hbar},
\ee
in their Compton frequencies, which is commonly known as the Bloch frequency. As usual, we have included a factor of $1+\beta$ to account for the redshift anomaly, can be either $\beta_{\rm II}$ or $\beta_{\rm III}$. This frequency can be observed as an oscillation of the atom's velocity expectation value over time.

Ivanov {\em et al.} \cite{Ivanov} used Strontium atoms ($^{88}$Sr) in an optical lattice of $\lambda/2$=266\,nm. Comparing the measured Bloch frequency with the expected one led to \cite{redshift}
\be
\beta=(4.0\pm6.0)\times 10^{-5}.
\ee
A similar measurement has been reported \cite{Clade} with rubidium ($^{87}$Rb) atoms and
$\lambda/2=394\,$nm leads to \cite{redshift}
\be
\beta =(3\pm1)\times 10^{-6}.
\ee
Meanwhile, preliminary data obtained by Guglielmo Tino (private communication and \cite{Alberti2010}) suggest that the accuracy of the Bloch oscillation measurement has been improved to several parts in $10^{-7}$.

We conclude that Pound-Rebka like atomic physics redshift measurements lead to limits of a few parts in $10^6$ at present. This is already an improvement by a factor of more than 10 compared to the best classical tests. Note that they are also particularly simple to interpret, as the atoms are stationary.

\subsection{``GP-A like" tests: atom interferometers}

\begin{figure}
\centering
\epsfig{file=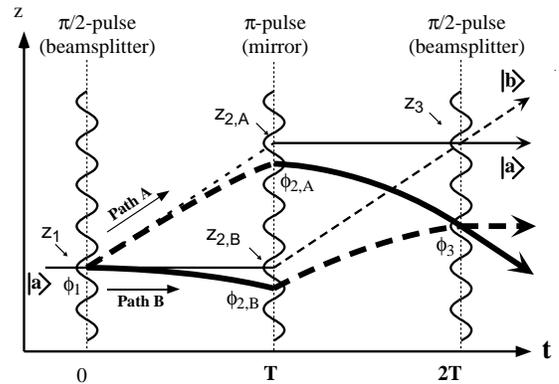,width=0.4\textwidth}
\caption{\label{MZ} Mach-Zehnder atom interferometer. The light straight lines indicate the trajectories without gravity, the bold lines the actual trajectories. A laser-cooled atom in a state $\left | a\right>$ is launched vertically upwards in a vacuum chamber. At time $t=0$, it encounters a first pulse of two counterpropagating laser pulses transfers having respective wavenumbers of $k_1$ and $k_2$. These pulses transfers the momentum $\hbar k$, where $k=k_1+k_2$ to the atom and change the internal state of the atom into $\left|b \right>$ (Fig. \ref{Raman}). The intensity and duration of the first laser pulse is adjusted such that this process happens with a probability of 50\%. As a result, the first laser pulse places the atom into a coherent superposition of two quantum states, which physically separate. The second pulse redirects the atom momentum so that the paths merge at the time of the third pulse.}
\end{figure}

\begin{figure}
\centering
\epsfig{file=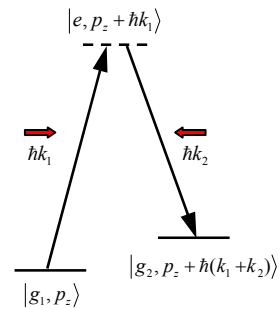,width=0.2\textwidth}
\caption{\label{Raman} Two-photon Raman beam splitter. An atom
in a quantum state $\left|g_1,p_z\right>$, moving upwards with momentum $p_z$, interacts with photons of two counter-propagating laser beams. The first one transfers the momentum $k_1$ and brings the atom into a virtual excited state $\left|e,p_z+\hbar k_1\right>$. The second laser beam stimulates the atom to emit a photon of momentum $k_2$, which transfers the atom to another hyperfine ground state $\left|g_2,p_z+\hbar (k_1+k_2)\right>$. With appropriate duration and intensity of the laser pulses, the process can have 50\% or 100\% probability, creating beam splitters or mirrors for atomic matter waves.}
\end{figure}

Just as GP-A used a large clock separation to obtain a higher sensitivity than Pound-Rebka tests, atom interferometers (Figs. \ref{MZ}, \ref{Raman}) obtain higher sensitivity than the Bloch oscillation tests by increasing the distance separating the atoms. At the same time, however, the atoms are moving. Thus, corrections for the special relativistic time dilation will have to be applied in order to extract the gravitational redshift. Since tests of the gravitational redshift necessarily operate outside the safe haven of the Einstein Equivalence Principle, the acceleration of free fall itself might be affected by the nonstandard physics. The trajectory therefore needs to be monitored to account for this.

The atom interferometer contains a built-in mechanism for monitoring the trajectories and applying this correction. To see this, we derive the phase difference between the interferometer arms using the standard procedures \cite{Young}. Since we do not assume prior knowledge of the acceleration of free fall, we will label it $g'$ and treat it as unknown. The specific value taken by $g'$ depends upon which of the three anomalous redshift phenomenologies scenarios is under test. The free evolution phase is derived by integrating the Compton frequency along the two paths 1,2 (Fig. \ref{MZ})
\bea\label{Ltype2}
\Delta \phi_{\rm free}&=&\int_1 \frac{mc^2}{\hbar} d\tau-\int_2\frac{mc^2}{\hbar}d\tau \nonumber \\ &=& -\om_C\int\left[(1+\beta)g\frac{z_1-z_2}{c^2}-\frac12 \frac{\dot z_1^2-\dot z_2^2}{c^2}\right]dt \nonumber\\ &=& (1+\beta)kgT^2-kg'T^2.
\eea
The first term, due to the redshift, contains $g$. It is given by the integral $\om_C (g/c^2)\int (\Delta U/c^2)dt,$ which depends on the vertical separation of the trajectories but not their acceleration of free fall. The second depends on $g'$, because it is given by how fast the atoms actually move.

Whenever a photon is absorbed by the atom, the photon's phase is added to the phase of the matter wave. For stimulated emission of a photon, the opposite occurs. The photon's phase $kz+\phi_0$ is a function of the location $z$ of the interaction occurs (where $\phi_0$ is its value at $z=0$). Summing up the photon's phases over all interactions (Fig. \ref{MZ}), we obtain
\be
\phi_{\rm i}=+kg'T^2.
\ee
This, again, contains the acceleration of free fall $g'$. The final result for the atom interferometer phase can be written as
\be\label{MZphase}
\Delta\phi=\underbrace{(1+\beta)kgT^2-kg'T^2}_{\Delta \phi_{\rm free}}+kg'T^2.
\ee
Thus, it turns out that the atom-light interaction phase cancels the one due to special relativistic time dilation no matter what the acceleration of free fall $g'$. This is analogous to measuring the trajectory and subtracting time dilation effects in classical redshift tests. This derivation can be made specific to each of the three classes of redshift anomalies described above by determining the expression for $g'$:
\begin{itemize}
\item In type I scenarios, $g'=(1+\beta)g$ as derived by extremizing the action Eq. (\ref{betalag}). Thus, $\Delta \phi_{\rm free}=0$ and $\Delta\phi=(1+\beta)kgT^2$. The parameter $\beta$, however, is not measurable. This becomes obvious when we express $\Delta\phi=kg'T^2$ and note that $g'$ is the acceleration of free fall as measured by the gravimeter.

\item For type II scenarios, $\beta$ is specific to the test particle T. We find $\Delta \phi_{\rm free}=0$ and $\Delta\phi=(1+\beta^{\rm T}-\beta^{\rm gravimeter})kgT^2$, where $\beta^{\rm gravimeter}$ is the parameter specific to the gravimeter. For a specific experiment, we may set it to zero by definition of coordinates.
\item For type III, $g=g'$ by definition, and $\Delta \phi=(1+\beta_{\rm III})kgT^2$, where $\beta_{\rm III}$ is now universal for all particles but will turn out to be a fixed function of the particle positions, momenta, and the gravitational acceleration. See Sec. \ref{typeIII} for a detailed model.

\end{itemize}

The most accurate data to date comes from an interferometer using caesium atoms in an atomic fountain \cite{Peters}. After correcting for a number of relatively small fundamental \cite{Petersmetrol,Dimopoulos} and systematic \cite{Peters,Petersmetrol} effects, we obtain the redshift parameter $\beta=z_{\rm meas}/z_0=(7\pm7)\times 10^{-9}$. In that experiment, the acceleration of gravity $g$ was measured using an FG-5 falling corner cube gravimeter. It basically measures the acceleration of free fall of a corner cube reflector which forms an optical interferometer (Fig. \ref{GenericTest}). 
Note that, in contrast to the atom interferometer, this is insensitive to the proper time experienced by the mirror, as the photons are interfered, not the wave packet (or density matrix) describing the mirror. 

\section{Description of matter wave tests in some test theories}

\subsection{Type II}\label{SME}

The standard model extension (SME) by Kosteleck\'y and co-workers provides us with a model for type II violations if taken to post-Newtonian order PNO(3). It is a general framework for describing Lorentz and CPT violation, starting from the Lagrangian of the standard model and adding all additional, Lorentz- and CPT violating, terms that can be formed from the standard model fields and Lorentz tensors. Recently, it was extended to incorporate gravity \cite{KosteleckyGravity,BaileyKostelecky,KosteleckyTassonPRL,KosteleckyTasson}.

To PNO(3), the relevant quantities are the pure-gravity sector coefficients $\bar s_{\mu\nu}$ as well as the matter sector coefficients $\alpha(\bar a^w_{\rm eff})_\mu$ and $(\bar c^w)_\mu$. The superscript $w$ can take the values $e,n,p$ indicating the electron, neutron, and proton, respectively. These Lorentz vectors and tensors that are defined in one particular inertial reference frame; it is conventional to adopt a sun-centered celestial equatorial reference frame \cite{BaileyKostelecky}. The Lorentz-violating properties of an object $B$ composed of $N^w$ of these particles can often be represented by effective coefficients
\bea\label{effectiveac}
(c^B)_{\mu\nu}&=&\frac{1}{m^B}\left(\sum_w N^w m^w (c^w)_{\mu\nu}+m'{}^{B}(c'{}^{B})_{\mu\nu}\right),\nonumber \\ (a_{\rm eff}^B)_\mu &=&\sum_w N^w (a_{\rm eff}^w)_\mu+(a'{}_{\rm eff}^{w})_\mu,
\eea
where $(a'{}_{\rm eff}^{w})_\mu$ and $(c'{}^{B})_{\mu\nu}$ indicate a possible Lorentz violating influence associated with the binding energy $m'{}^{B}$.

The acceleration of free fall of a light test mass $m^{\rm T}$ is given by \cite{KosteleckyTasson}, Eq. (138) or (229)
\bea
\tilde g &=& g_{\rm S}\left(1+\frac{2\alpha}{m^{\rm T}}(\bar a_{\rm eff}^{\rm T})_0-\frac 23 (\bar c^{\rm T})_{00}\right),\nonumber \\
g_{\rm S}&=& g\left(1+\frac{2\alpha}{m^{\rm S}}(\bar a_{\rm eff}^{\rm S})_0
+(\bar c^{\rm S})_{00}+\frac 53 \bar s_{00}\right)
\eea
where $g$ is the acceleration of free fall when the SME-coefficients vanish and  $g_{\rm S}$ is a function of the SME-coefficients of the source and the gravitational sector. By imposing the condition
\be\label{UFFSME}
\alpha(\bar a^w_{\rm eff})_0=\tfrac13 m^w(\bar c^w)_{00}
\ee
and neglecting the binding energy in Eq. (\ref{effectiveac}), we obtain the ``isotropic parachute model" \cite{KosteleckyTasson} in which UFF is valid.



To determine the gravitational shift in the Compton frequency $m c^2/\hbar$ of matter waves, the dependence of the rest mass $m$ on the metric $g_{\mu\nu}=\eta_{\mu\nu}+h_{\mu\nu}$ has to be found. This can be accomplished by collecting all terms that are independent of the particle momentum in the Hamiltonian $H_{\rm NR}$ given in Ref. \cite{KosteleckyTasson}. At PNO(3), we find
\bea
H_{\rm NR}&=&\tfrac12 m^{\rm T}v^2\left(1+\tfrac53 (c^{\rm T})_{00}+\tfrac32 h_{00}
\right)\nonumber \\ && -\tfrac12 m^{\rm T}h_{00}\left(1+\frac{2\alpha}{m^{\rm T}}(\bar a_{\rm eff}^{\rm T})_0+(c^{\rm T})_{00}\right)
\eea
If Eq. (\ref{UFFSME}) and, thus, UFF, hold, this hamiltonian becomes
\bea
H_{\rm NR}&=&\tfrac12 m^{\rm T}v^2\left(1+\tfrac53 (c^{\rm T})_{00}+\tfrac32 h_{00}
\right)\nonumber \\ && -\tfrac12 m^{\rm T}h_{00}\left(1+\tfrac53 (c^{\rm T})_{00}\right).
\eea
Both the kinetic and potential terms carry a constant factor of $1+\tfrac 53 (c^{\rm T})_{00}$. This can be eliminated from the equations of motion by rescaling the particle masses to become $m^{\rm T}{}'=m^{\rm T}(1+\tfrac53 (c^{\rm T})_{00})$, an unmeasurable scaling of the unit of mass.  Thus, to PNO(3), there is no type III redshift violation in the SME.

Without Eq. (\ref{UFFSME}), however, the kinetic term is scaled differently than the potential term.  We can still redefine the particle mass to be $m^{\rm T}{}'=m^{\rm T}(1+\tfrac53 (c^{\rm T})_{00})$, to put the kinetic term in the form expected by Eq. (\ref{Ltype2}), but then we find that the potential term is scaled by
\be
-\tfrac 12 m^{\rm T}{}' h_{00}\left[1+\frac{2\alpha}{m^{\rm T}{}'}(\bar a^{\rm T}_{\rm eff})_0-\tfrac23 (\bar c^{\rm T})_{00}\right].
\ee
This yields Eq. (\ref{Ltype2}) with
\be
\beta^{\rm T} = \tfrac 23\left[\frac{3\alpha}{m^{\rm T}{}'}(\bar a^{\rm T}_{\rm eff})_0 -(\bar c^{\rm T})_{00}\right].
\ee
The superscript T at the redshift parameter $\beta$ indicates the nature of the particle. From the matter-wave experiments described above, we obtain limits of
\bea
\beta^{88\rm Sr} &=&(4\pm 6)\times 10^{-5},\nonumber \\
\beta^{87 \rm Rb} &=&(3\pm 1) \times 10^{-6}
\eea
from Bloch oscillation experiments, and
\be
\beta^{133\rm Cs}=(7\pm7)\times 10^{-9}
\ee
from the atom interferometer.

\subsection{Type III}\label{typeIII}

The Standard Model Extension also provides a model for type III violations if it is taken to PNO(4). We will present such a model in a separate publication, because taking the SME to PNO(4) is quite challenging given the nonlinearity of General Relativity plays a substantial role at this order. In this section, we offer a toy model \cite{MikeCPT} in which particle masses anomalously vary with the gravitational potential. The anomaly is represented by a single parameter $\beta_{\rm III}^0$ (which represents a combination of SME parameters). This generates gravitational redshift anomalies unconstrained by existing tests of UFF or LLI. The effects predicted by this toy model are qualitatively similar to those generated by the SME at PNO(4), although it should also be noted that similar effects exist in the fully Lorentz covariant theory, due to the nonlinearity of general relativity.

We consider a model where particles' masses anomalously vary in their local gravitational potential
\begin{equation}
L=mc^{2}+mc^{2}\left(1+\beta_{\rm III}^0\frac{gz}{c^{2}}\right)\left(\frac{gz}{c^{2}}-\frac{\dot{z}^{2}}{2c^{2}}\right)\label{eq:lagIII}.
\end{equation}
This model includes terms resembling those appearing at higher order in the expansion of the Schwarzschild metric. ($\beta_{\rm III}^0\neq 0$ may be understood as several anomalous terms in the expansion.  Higher order gradients and GR must be considered when experiments are sensitive to $\beta_{\rm III}^0 \sim 6$.) This Lagrangian modifies point-particle kinematics: the local acceleration of free-fall is, to first order in $\beta_{\rm III}^0$,
\begin{equation}
\ddot{z}=-g\left(1+\beta_{\rm III}^0\left[\frac{gz}{c^{2}}+\frac{\dot{z}^{2}}{2c^{2}}\right]\right),
\end{equation}
and is the same for all objects.  Expanding to $\mathcal O\left(T^{2}\right)$, where $T$ is the pulse separation time in the interferometer, and assuming that photon momenta scale like the masses with height (dropping this assumption changes the second term given below by a factor of $3$), the total  AI phase is
\begin{equation}
\Delta\phi = kgT^{2}-\beta_{\rm III}kgT^{2}-\beta_{\rm III}^0\frac{v_{0}v_{r}}{c^{2}}kgT^{2},\label{eq:typeIIIphase}
\end{equation}
where
\be
\beta_{\rm III}\equiv \beta_{\rm III}^0\frac 12 \frac{v_0^2}{c^2}
\ee
General relativity, due to its nonlinearity, predicts an interferometer phase similar to the second term in \refeq{eq:typeIIIphase} (with $\beta_{\rm III}^0=10$) \cite{Dimopoulos}. Thus the type III violations are quantitative modifications of effects that already exist in general relativity.

The most sensitive tests of the gravitational redshift in AIs are precise to 7 ppb in $\beta_{\rm III}$ \cite{redshift}, which in light of the small launch velocity ($v_{0}=1.53$ m/s), translates to a bound of $|\beta_{\rm III}^0|\leq 5\times 10^{8}$. This sensitivity is not unique to the AI.  Moving Pound-Rebka-Snider tests with $\vec{v}\cdot{\vec{g}}/|\vec{g}|=v_{0}\ll c$ observe
\begin{equation}
\nu_{2,1}=\nu_{0}\left(1+\frac{gh}{c^{2}}\left[1-\beta_{\rm III}^0 \frac{v_{0}}{c}\left(1-\frac{v_{0}}{c}\right)\right]\right),
\end{equation}
where we ignore higher order terms in the Schwarzschild expansion which become relevant when $\beta_{\rm III}^0$ becomes $1$ or lower.  Motion of the Earth in the Sun's gravitational potential would permit Earth-stationary Pound-Rebka-Snider experiments with $1\%$ accuracy to bound $|\beta_{\rm III}^0|<10^{8}$ by repeated measurements made while the Earth moves between orbital perihelions. However, such an experiment has yet to be performed.

\paragraph{Existing limits}

It is perhaps surprising that effects as large as $|\beta_{\rm III}^0|=10^8$ are not yet ruled out by direct experimental measurement. This is because of the suppression by gravity as well as velocity, and because no experiment has yet been optimized to detect the effect.

A generalization of this model is constrained by solar system tests of GR.  One possible generalization of \refeq{eq:lagIII} to a two-body problem might be
\begin{equation}
L=\left[\frac{1}{2}\left(M\dot{\vec{R}}^{\,2}+\mu\dot{\vec{r}}^{\,2}\right)
+\frac{GM\mu}{r}\right]\left(1+\beta_{\rm III}^0\frac{GM}{rc^{2}}\right),\label{eq:3d}
\end{equation}
where $M=m_{1}+m_{2}$ is the total mass, $\mu=m_{1}m_{2}/M$ is the reduced mass, and $\vec{R}$ and $\vec{r}$ are the center of mass and relative position vectors.
In the limit $\mu\ll M$, and $z=r-r_{\oplus}\ll r_{\oplus}$, with $r_{\oplus}$ the Earth's radius, we recover the one-body Lagrangian \refeq{eq:lagIII}.  Eq. \refeq{eq:3d} causes anomalous periapsis precession, $\bar{\dot{\omega}}=\beta_{\rm III}^0\bar{\dot{\omega}}_{\rm GR}$, where $\bar{\dot{\omega}}_{\rm GR}$ is the GR precession.  The perihelion precession of Mercury fractionally differs from that predicted by GR by less than $1\times 10^{-3}$~\cite{Will:2006}, implying $|\beta_{\rm III}^0|\leq 1\times10^{-3}$.  This constraint may not apply to terrestrial AI tests, as $\beta_{\rm III}^0$ might vanish at large $r$ in another model.



\section{Clock-comparison versus matter wave tests}

In a type II model, clock comparison redshift tests have been studied in \cite{KosteleckyTasson}. The gravitational modifications in the clock's hamiltonian modify the redshift according to $\Delta \nu/\nu=(\tilde gz/c^2) (1+\xi_{\rm clock}),$ where $\xi_{\rm clock}$ is a clock-dependent parameter that can be calculated if the hamiltonian of the clock is known. For the transition frequencies in a Bohr-model hydrogen atom, for example \cite{KosteleckyTasson}
\be
\xi_{\rm H}=-\frac 23 \left(\frac{m^p}{m^{\rm H}}(\bar c^e)_{00}+\frac{m^e}{m^{\rm H}}(\bar c^p)_{00}\right).
\ee
This shows that in type II models, atomic clock redshift tests are predominantly sensitive to the electron coefficients $\bar c^e$ and have a suppressed sensitivity towards proton and neutron terms. The reverse is true for matter wave tests. Both, however, compete with tests of UFF in this scenario. As mentioned before,  classical redshift tests would need to reach $2\times 10^{-7}$ before they can become competitive.

In type III systems, the redshift violation parameter is independent of the nature of the clock by necessity, because otherwise UFF would be violated. Hence, there is no difference whatsoever between matter wave redshift tests and clock comparisons. Establishing the existence or non-existence of type III effects is the most important task of redshift experiments, as this has not yet been done by tests of UFF.

As this comparison shows, the criticism that atom interferometers do not measure the gravitational redshift \cite{comment} is baseless \cite{reply}: both experiments measure similar SME coefficients in type II scenarios, and precisely the same in type III scenarios.

\section{Geodesic explorer: atom interferometric redshift test in space}

We here propose a mission with an atom interferometer in a sounding rocket or orbiting satellite to test the gravitational redshift \cite{HolgerCPT}. It will improve the accuracy of present redshift experiments by a factor of $10^5$ for type II effects or $10^{12}$ for type III effects. Operation in space is crucial, as it converts the measurement into a null measurement and provides an increase of the boost factor $v^2/c^2$ by about 7 orders of magnitude. Since type II violations are equivalent to violations of UFF, the Explorer will also bound violations of UFF at an accuracy 10 times better than current bounds. Moreover, it will test UFF for a quantum object versus a classical object. The Explorer will also provide the best bounds on a so far unbounded class of parameters of the SME and can be configured to perform numerous other tasks, such as geodesy, measuring the fine structure constant \cite{SCI,Biraben} and measuring the Lense-Thirring effect.

The signal in ground-based atom interferometry measurements of the gravitational redshift is given by the contributions of the redshift $\phi_{\rm r}$, time dilation $\phi_{\rm t}$, and laser-atom interaction $\phi_{\rm i},$
\be
\phi_{\rm ground}=\phi_{\rm r}+\phi_{\rm t}+ \phi_{\rm i}=(1+\beta) kgT^2.
\ee
See above for the specific values of the three terms for violations of type II and III. For a freely falling interferometer, however, $\phi_{\rm i}=0$ while the other two terms are unchanged (as the atoms are in free fall even in the ground-based instrument). Such a freely falling interferometer can be realized on a microgravity platform. In this case, the phase is given by
\be
\phi_{\rm space}=\phi_{\rm r}+\phi_{\rm t}=\beta kgT^2.
\ee
Thus, space-based operation converts the redshift measurement into a null measurement: the ordinary effects of gravity vanish, whereas a redshift anomaly will still produce a signal. This was realized in the context of clock comparison tests by Krisher \cite{Krisher1995}. It removes the major limiting uncertainty, the one in the measurement of local gravity. A redshift violation of $\beta\sim 10^{-14}$ can produce several microradian of phase shift between the interferometer arms, which can be measured at the shot noise level within one week of integration time.

While the Geodesic Explorer might thus improve the sensitivity to type II effects by 5 orders of magnitude, an additional 7 orders can be gained towards type III effects: The leading effects of $\beta_{\rm III}^0$ in the Earth's gravitational field [see Eq. (\ref{eq:typeIIIphase})] are suppressed by a factor of $g v^2/c^4$, where $v$ is the characteristic velocity of the apparatus along the gravitational potential gradient. This is because these effects are PNO(4) effects, proportional to $1/c^4$. These terms include ones proportional to $(\Delta z)^2 g^2/c^4$ and $\Delta z g v^2/c^4$, where $\Delta z$ is the distance of the two clocks and $v$ a typical velocity of the apparatus. Operating an AI redshift test on a sounding rocket~\cite{Reasenberg:2010} or an Earth satellite on a highly elliptical orbit will amplify the $(v/c)^{2}$ term by a factor of $\sim 10^{7}$ relative to the stationary test~\cite{Vessot}.  Such a test could have a sensitivity to $\beta_{\rm III}^0\sim 1\times 10^{-4}$, 12 orders of magnitude better than the present AI bound (see above). This would represent one of the first experimental tests of GR at PNO(4). Note that such a test would also be sensitive to higher order GR effects not accounted for here~\cite{Dimopoulos}.

To remove systematic effects, it is important to note that the frame of reference of the atom interferometer can be defined by using a retroreflection mirror to make the counterpropagating laser beams. It is the motion of this mirror that defines the frame of reference to which the atom interferometer signal is referred. Vibrations and residual accelerations of the rest of the setup are then unimportant. One experimental option is to float the mirror inside the spacecraft and servo the position of the spacecraft relative to the mirror drag-free control. The precision of the drag-free control can be moderate, as it is only needed to keep the mirror's position relative to the spacecraft within acceptable limits that may easily be a few mm. Alternatively, operation without drag-free control is possible if the mirror is floating during interferometer operation, but repositioned periodically.

Table \ref{GExplorer} shows some gravitational and systematic effects in a spaceborne redshift experiment. The largest uncertainties besides residual acceleration is the gravity gradient. The gravity gradient error can be reduced effectively by proper spacecraft design similar to the approach used for the LISA and STEP \cite{Step} missions. Magnetic fields will have to be suppressed to the microgauss level, which can be done by shielding, or by a double-interferometer wherein half the atoms operate as a magnetometer. The systematic effects are compatible with reaching a sensitivity of $10^{-14}$ in the redshift parameter, a 100,000 fold improvement relative to the best Earth-based experiment.

Unlike other proposed space-based redshift tests \cite{EGE,ACES}, the Geodesic Explorer will not compare a clock in orbit to a clock on ground but rather two clocks at different elevations but both in orbit. Thus, the gravitational potential of the spaceborne clock versus Earth does not have to be known to the full accuracy of the experiment. This, and the extreme oscillation frequency of matter wave clocks, allows us to reach an accuracy of $10^{-14}$ compared to $10^{-8}$ for the most ambitious clock-comparison test \cite{EGE}. In type II scenarios, the $10^{-14}$ bound from the AI provides an equivalent bound of $10^{-8}$ on the electromagnetic binding energy $\beta^{\rm em}$. Note that if a conventional clock comparison test reached an accuracy of $10^{-9}$, its sensitivity to $\beta^{\rm em}$ would be superior to that of an AI test with $10^{-14}$ accuracy, although the conventional clock experiment would be far less sensitive to the $\beta^p$ and $\beta^n$ terms associated with the proton and neutron.

The Explorer will also measure the $a-$type coefficients for Lorentz violation in the fermion sector of the SME. These coefficients are unconstrained so far. Thus, there exists a possibility of large Lorentz violations that would so far have been undetected \cite{KosteleckyTassonPRL}. The geodesic explorer could bound these coefficients at a level of at least $10^{-14}$ and better for some of the coefficients: The satellite's motion would cause a specific time modulation of the a-coefficients. in the satellite frame. By analyzing the signal over several orbits and extracting the modulation amplitudes, suppression of systematic effects can be reached.

\begin{table*}
\caption{\label{GExplorer} Gravitational effects in a space-borne atom interferometer. We assume a redshift anomaly of $\beta=10^{-14}$, Cs atoms in the $m_F=0$ state, a wavelength of 852\,nm, an orbit of 340\,km, $n=10$-photon beam splitters, same internal states, a pulse separation time $T=10\,$s, an initial atom velocity $v_0=1\,\mu$m/s, and a residual gravitational acceleration $g'=10^{-14}\,$m/s$^2$. `DK$n$' means line $n$ in Table 1 of Ref. \cite{Dimopoulos}.}
\begin{tabular}{cc}\hline\hline
Effect & Phase (rad) \\ \hline
Redshift anomaly & $6.5\times 10^{-5} (\beta/10^{-14}) (n/10) (T/10{\rm ~s})^2$ \\
Residual acceleration & $7.3\times 10^{-5} (n/10) (T/10 {\rm ~s})^2 (g'/10^{-14}{\rm ~m s}^{-2})$ \\
Gravity gradient & $-1.8\times 10^{-1} (n/10) (T/10{\rm ~s})^3 (v_0/\mu{\rm ~m s}^{-1})$ \\
Gravity gradient & $-1.1\times 10^2 (n/10) (T/10{\rm ~s})^4 (g'/10^{-5}{\rm ~m s}^{-2})$ \\
Finite speed of light & $-7.3\times 10^{-8} (n/10) (T/10{\rm ~s})^3$ \\ Doppler effect & $-7.4\times 10^{-10} (n/10) (T/10 {\rm ~s})^2 (v_0/\mu{\rm ~m s}^{-1}) (g'/10^{-5}{\rm ~m s}^{-2})$ \\
First gradient recoil & $-3.2\times 10^3 (n/10)^2 (T/10 {\rm ~s})^3$ \\ DK9 & $-1.3\times 10^{-5} (n/10)^2 (T/10{\rm ~s})^2 (g'/10^{-5}{\rm ~m s}^{-2})$ \\
Raman splitting DK13$^a$ & $-1.3\times 10^{-13} (n/10) (T/10{\rm ~s})^2 (f_{\rm HFS}/9 {\rm ~GHz}) (g'/10^{-5}{\rm ~m s}^{-2})$ \\
$g^3$ DK17 & $-5.7\times 10^{-20} (n/10) (T/10{\rm ~s})^4 (g'/10^{-5}{\rm ~m s}^{-2})$\\
Gravity gradient DK25 & $-7.2\times 10^{-9} (n/10)^2 (T/10{\rm ~s})^4 (g'/10^{-5}{\rm ~m s}^{-2})$ \\
Raman splitting DK26$^a$ & $2.3\times 10^{-11} (n/10) (T/10{\rm ~s})^2 (f_{\rm HFS}/9 {\rm ~GHz}) (g'/10^{-5}{\rm ~m s}^{-2})$  \\
Shot noise & $10^{-6}({\rm ~flux}/10^6{\rm ~s}^{-1})^{-1/2}({\rm time}/{\rm week})^{-1/2}$ \\
Magnetic fields & $1.6\times 10^{-2}(n/10)(T/10{\rm ~s})^2(B{\rm d} B/{\rm d} z/({\rm mG})^2{\rm ~m}^{-1})$\\
Cold collisions & $10^{-4}({\rm ~density}/10^6{\rm ~cm}^{-3})({\rm balance}/0.1) $ \\ \hline \hline
\end{tabular}
\\
{\footnotesize $^a$ zero for same internal states.}
\end{table*}

\acknowledgments We thank Sven Herrmann, Mark Kasevich, Juna Kollmeier, Alan Kosteleck\'y, Silke Ospelkaus, Jay Tasson, and Guglielmo Tino for numerous important discussions. We also thank Peter Hannaford and all the Organizers of the 22. International Conference on Atomic Physics held in 2010 in Cairns, Australia. Support by the David and Lucile Packard Foundation, the Alfred P. Sloan Foundation, and grant No. 60NANB9D9169 of the National Institute of Standards and Technology is acknowledged.
\appendix

\section{Equivalence to the Schr\"odinger equation}

From Eq. (\ref{Comptonphase}), it is clear that many effects in quantum mechanics are connected to the gravitational redshift and special relativistic time dilation. They can, therefore, be employed in testing general relativity. Since most experiments are nonrelativistic, the path integral is usually unnecessarily complicated. It is therefore interesting to develop the above ideas into a more familiar form that is directly applicable to nonrelativistic quantum mechanics. Here, we will show that the interpretation of atom interferometry as redshift tests is mathematically equivalent to the Schr\"odinger equation of an atom in a gravitational field. We shall follow the approach of Feynman \cite{Feynman1948}. This approach is, thus, not fundamentally new. However, there is pleasure in viewing familiar things from a new point of view. We start by using a post-Newtonian approximation
\bea
S&=&\int mc^2 \sqrt{1+h_{\mu\nu}u^\mu u^\nu}d\lambda \nonumber \\
&\approx & \int mc^2 \left(1+\tfrac 12 h_{00}+ h_{0j}\frac{u^j}{c}-\tfrac12(\delta_{jk}-h_{jk})\frac{u^j}{c}\frac{u^k}{c} \right)dt\nonumber\\
\eea
where $u^j$ is the usual 3-velocity. We replaced the parameter $\lambda$ by the coordinate time $t$. We now compute the path integral for an infinitesimal time interval $\epsilon$, during which the integrand can be treated as constant. We denote $\vec \xi=\vec x_B-\vec x_A$. For an infinitesimal $\epsilon,$ $\vec v=\vec \xi/\epsilon$, so
\bea
\psi(t+\epsilon,\vec x_A)&=&N\int d^3\xi\, \psi(t,\vec x_A-\vec \xi) \nonumber \\ &&\times \exp\left(i \frac{mc^2\eps}{\hbar}\left(1+\tfrac 12 h_{00}\right)\eps\right) \nonumber \\ &&\times \exp\left[-\frac{1}{2} A_{jk}\xi^j\xi^k + B_j\xi^j \right]
\eea
where $N$ is a normalization factor and
\be A_{jk}\equiv -\frac{im}{\hbar \eps} (\delta_{jk}+h_{jk}), \quad B_j\equiv  \frac{im c }{\hbar}h_{0j}.
\ee.
We can expand in powers of $\eps, \xi$:
\bea
\psi(t+\epsilon,\vec x_A)&=&N\int  \left(\psi-\xi^j \partial_j \psi+\tfrac12 \xi^j\xi^k \partial_j\partial_k \psi\right) \nonumber \\  && \times \left(1+i \frac{mc^2\eps}{\hbar}(1+\tfrac12 h_{00})\right)\nonumber \\ &&\times \exp\left[-\frac{1}{2} A_{jk}\xi^j\xi^k + B_j\xi^j\right]d^3\xi
\eea
where $\psi\equiv \psi(t,\vec x_A)$. We compute the integrals \cite{Zee}
\be
\int
e^{ -\frac{1}{2} A_{jk}\xi^j\xi^k + B_j\xi^j}d^3\xi =\frac{(2\pi)^{3/2}}{\sqrt{\det A}}e^{-\frac12 B_j(A^{-1})_{jk}B_k},
\ee
where $\det A$ is the determinant of $A$ and $A^{-1}\approx -\frac{\hbar \eps} {im}(\delta_{jk}-h_{jk})$ is the inverse matrix. We obtain
\bea
\psi(t+\epsilon,\vec x_A)&=&N\frac{(2\pi)^{3/2}}{\sqrt{\det A}} \left[\left(1+i \frac{mc^2\eps}{\hbar}(1+\tfrac12 h_{00})\right)\psi \right. \nonumber \\ && \left.
-(\partial_j\psi)\frac{\partial}{\partial B_j}
+\frac 12 (\partial_j\partial_k\psi)\frac{\partial}{\partial B_j} \frac{\partial}{\partial B_k}  \right] \nonumber \\ && \times \exp\left(\frac12 B_jB_k(A^{-1})_{jk}\right).
\eea
The normalization factor is determined from the fact that $\psi(t+\epsilon,\vec x_A)$ must approach $\psi(t,\vec x_A)$ for $\eps\rightarrow 0$. We find
\be
N= \frac{\sqrt{\det A}}{(2\pi)^{3/2}}e^{-\frac12 B_jB_k(A^{-1})_{jk}}.
\ee
We carrying out the derivatives with respect to $B_j$ and inserting $B_j$ and $(A^{-1})_{jk}$.
Working in post-Newtonian order 3, we can neglect $h_{jk}$ and neglect terms proportional to $\eps^2$.
This leads to a Schr\"odinger equation

\be\label{Schrodinger}
i\hbar \frac{d}{dt} \psi =- mc^2\tfrac12 h_{00}\psi
- \frac{\hbar^2}{2m} \left(\vec \nabla-m\vec h \right)^2 \psi
\ee
where we have substituted $\psi\rightarrow e^{-i \om_C t}\psi$. The 3-vector $\vec h$ is defined by $h_j\equiv (i c/\hbar)h_{0j}$. We neglected a term proportional to $h_{0j}h_{0j}$ and one proportional to $h_{0j,j}$.

From the path integral approach, the usual commutation relations can also be derived \cite{Feynman1948}. This shows that quantum mechanics is a description of waves oscillating at the Compton frequency that explore all possible paths through curved spacetime.


\begin{references}
\bibitem{EEPremark} It is customary to define three ingredients of the Einstein Equivalence Principle \cite{TEGP}: (i) Local Lorentz invariance (LLI) refers to the frame-independence of non-gravitational physics. (ii) Universality of free fall (UFF) states the universality of the motion of co-located test particles, independent of the nature of the particle; (iii) Local position invariance (LPI) summarizes all further gravitational physics but the most important LPI tests are measurements of the gravitational redshift.
\end{references}
\end{document}